\documentclass[a4paper]{jpconf}
\usepackage{graphicx}
\begin{document}
\title{Three dimensional solar anisotropy of galactic cosmic rays near the recent solar minimum 23/24}

\author{R. Modzelewska$^{1}$, M.V. Alania$^{1,2}$}

\address{$^{1}$Institute of Mathematics and Physics, Siedlce University, Poland.}
\address{$^{2}$Institute of Geophysics, Tbilisi State University, Tbilisi, Georgia.}
\ead{renatam@uph.edu.pl, alania@uph.edu.pl,}

\begin{abstract}

Three dimensional (3D) galactic cosmic ray (GCR) anisotropy has been studied for 2006-2012. The GCR anisotropy, both in the ecliptic plane and in polar direction, were obtained based on the neutron monitors (NMs) and Nagoya muon telescopes (MT) data. We analyze two dimensional (2D) GCR anisotropy in the ecliptic plane and north-south anisotropy normal to the ecliptic plane. We reveal quasi-periodicities - the annual and 27-days waves in the GCR anisotropy in 2006-2012. We investigate the relationship of the 27-day variation of the GCR anisotropy in the ecliptic plane and in the polar direction with the parameters of solar activity and solar wind.

\end{abstract}

\section{Introduction}

Diurnal variation, being the measure of the galactic cosmic ray (GCR) anisotropy, arises from the Earth's rotation as a ground-based detector's viewing direction  through the sky during each 24-hours period. Three-dimensional (3D) anisotropy vector is determined by  the distribution of the stream of cosmic rays in the 3D heliosphere. An  average anisotropy vector has been explained based on the diffusion-convection theory of GCR modulation in the heliosphere \cite{Ahluwalia62}-\cite{Parker64}. That is a consequence of the equilibrium established between the radial convection by solar wind and the inward diffusion of GCR particles along the interplanetary magnetic field (IMF) owing to the radial gradient of GCR.

Scientific staff of IZMIRAN's  cosmic ray laboratory\\
(http://helios.izmiran.troitsk.ru/cosray/main.htm), have calculated the components  $Ar$, $Af$, and $A_{\theta}$ of the 3D anisotropy by Global Spectrographic Method (GSM) \cite{Belov05}-\cite{Krymsky67} based on hourly data from all operating neutron monitors (NMs).  Unfortunately, the derivation of the $A_{\theta}$   component is possible  with an accuracy up  to  constant, and so,  a value of the north-south anisotropy $A_{\theta}$   obtained by GSM method is not accurate (http://cr20.izmiran.rssi.ru/AnisotropyCR/index.php). That is, sorrowfully, a deficiency  of the GSM related to the nature of NMs data.

In general determination of the two dimensional (2D) anisotropy in the ecliptic plane is feasible based on establishing the radial $Ar$ and tangential $Af$ components by the harmonic analysis method for an individual detector (e.g., NM or Muon Telescope (MT)). Alania et al., \cite{Alania08} demonstrated  that the 2D GCR anisotropy calculated by the radial and tangential components determined using GSM basically do not differ from the values of anisotropy found by harmonic analysis method for individual NM with cut-off rigidities $< 5$ GV near the solar minimum epoch.

Swinson \cite{Swinson69} first suggested that the north-south anisotropy could be related to the cosmic ray flow caused by a positive heliocentric radial density gradient $\mathbf{Gr}$ of cosmic rays and the $\mathbf{By}$ component of the interplanetary magnetic field (IMF) $\mathbf{B}$, as expressed by the vector product, $\mathbf{By}$ x $\mathbf{Gr}$. Since the average  $\mathbf{By}$ and the $\mathbf{Gr}$ lie in the ecliptic plane, the direction of the vector product $\mathbf{By}$ x $\mathbf{Gr}$  (or the north-south anisotropy) is expected to be perpendicular to the ecliptic plane and to reverse direction with changes of the magnetic field direction. It is directed upward  when $\mathbf{By}$ is positive (IMF away from the Sun-$A>0$) and vice versa when $\mathbf{By}$ is negative (IMF toward the Sun-$A<0$).

\begin{figure}[tbp]
  \begin{center}
\includegraphics[width=0.9\hsize]{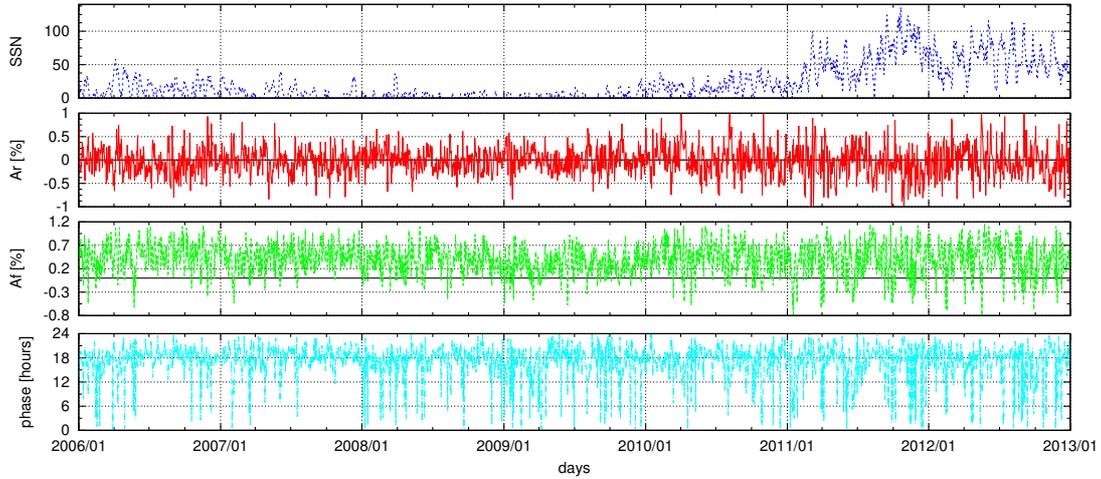}
\end{center}
\caption{\label{fig:Aniz} Daily data of the sunspot number $SSN$, radial $Ar$ and tangential $Af$ components and phase of the 2D GCR anisotropy for Oulu NM in 2006-2012. }
\end{figure}

\begin{figure}[tbp]
  \begin{center}
\includegraphics[width=0.9\hsize]{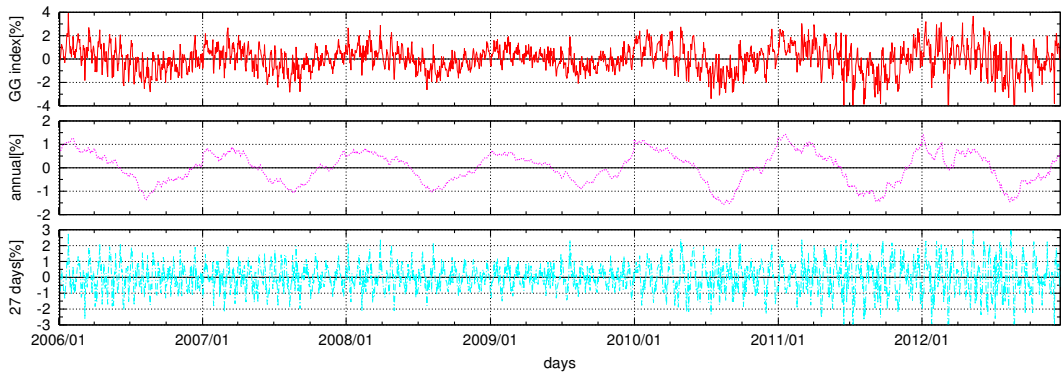}
\end{center}
\caption{\label{fig:GG} Temporal changes of the (top) daily $GG$ index for 2006-2012, (midle) annual trend in $GG$ index and (bottom) daily $GG$ index with excluded annual trend.}
\end{figure}

Papers \cite{DP77}- \cite{Bieber93} studied the cosmic ray anisotropy vector in three dimensions. They have determined north-south anisotropy using data from polar located NMs. They concluded that a magnitude of the north-south anisotropy varies with
$\sim 11$-years period and there is not any dependence on solar magnetic polarity.

\begin{figure}[tbp]
  \begin{center}
\includegraphics[width=0.7\hsize]{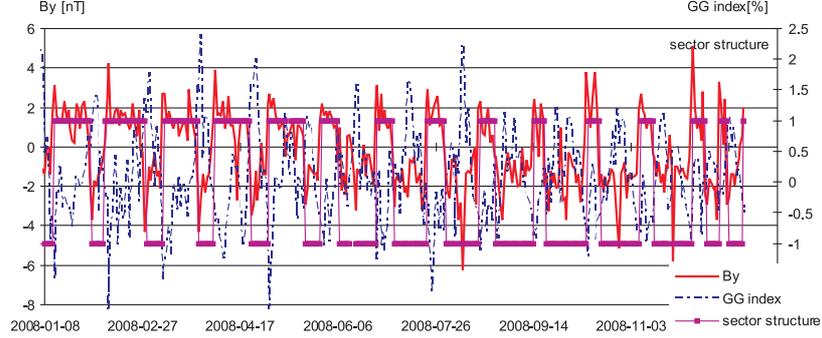}
\end{center}
\caption{\label{fig:2008} Daily $GG$ index, $By$ component and the sector structure of the IMF in 2008 }
\end{figure}

An alternative method to study north-south anisotropy is proposed by Mori and Nagashima \cite{MN79}. They have introduced an index $GG$ calculated from Nagoya MT data as:
\begin{eqnarray}
\label{GG}
GG = \frac{1}{2}[(49N-49S) + (49N-49E)]
\end{eqnarray}

The terms represent the counting rate differences for telescopes pointing in North, South, and East directions at $49^{0}$ zenith angle. The $GG$ index is free of noise in isotropic intensity caused by Forbush decreases, periodic variations, atmospheric temperature effects, and geomagnetic cut-offs. $GG$ index mainly reflects change of difference between the intensity from north polar direction and that from north and  parallel to the equatorial plane directions. Counting rate differences of the north-south and north-equatorial could  not be precisely contained  the same type of information. In spite, a $GG$  is accepted by cosmic ray community as a good alternative index to study the north-south (or north-equatorial) asymmetry of the cosmic ray flux e.g., \cite{Munakata14}.

The present paper investigates the annual and 27-day waves of the  GCR anisotropy based on ground observations of secondary cosmic rays by NMs and MTs near the recent solar minimum 2006-2012. We investigate the relationship of the 27-day variation of the GCR anisotropy in the ecliptic plane and in the polar direction with parameters of solar activity and solar wind.

\section{Data}
Figure ~\ref{fig:Aniz}  presents the temporal changes of the daily sunspot numbers (SSN),  daily radial $Ar$ and tangential $Af$ components,  and phase of the 2D GCR anisotropy obtained by Oulu NM for 2006-2012. Figure ~\ref{fig:Aniz}  shows that components of the daily ecliptic plane GCR anisotropy obtained from hourly data of NM experience large dispersions during the Sun's rotation period. However, one can recognize that  the average daily radial component $Ar$ oscillate near zero, which is caused by a sign dependent drift effect in well established sector structure of the IMF; the tangential component $Af$ ($\sim 0.4\%$) represents diffusive corotational anisotropy and phase (time of maximum intensity) indicates that GCR stream is directed averagely from $\sim 18h$.

Figure ~\ref{fig:GG}  (top panel) presents daily data of the $GG$ index for 2006-2012. Figure ~\ref{fig:GG}  shows that a clear annual trend is seen in changes of the  $GG$ index.

Our aim is to study  both annual wave and 27-day variation in $GG$ index. We reveal annual trend by smoothing daily data over 27 days (Fig. ~\ref{fig:GG}  middle panel). Finally for studying the 27-day variation of the $GG$ index we exclude the annual trend and present detrended daily $GG$ in the bottom panel of Fig.~\ref{fig:GG}.

As far the formulation of the north-south anisotropy is based on the drift model, it has been established that there is a good correlation between $GG$ index and the polarity of the IMF during each solar rotation \cite{MN79}. Consequently the $GG$ index being the measure of the north-south asymmetry is inversely related with $By$ component of the IMF. It is also well pronounced in the recent solar minimum 23/24. As an example Fig. ~\ref{fig:2008}  presents daily $GG$ index, $By$ component and the sector structure of the IMF in 2008.

   \begin{figure}[tbp]
  \begin{center}
 \includegraphics[width=1.0\linewidth]{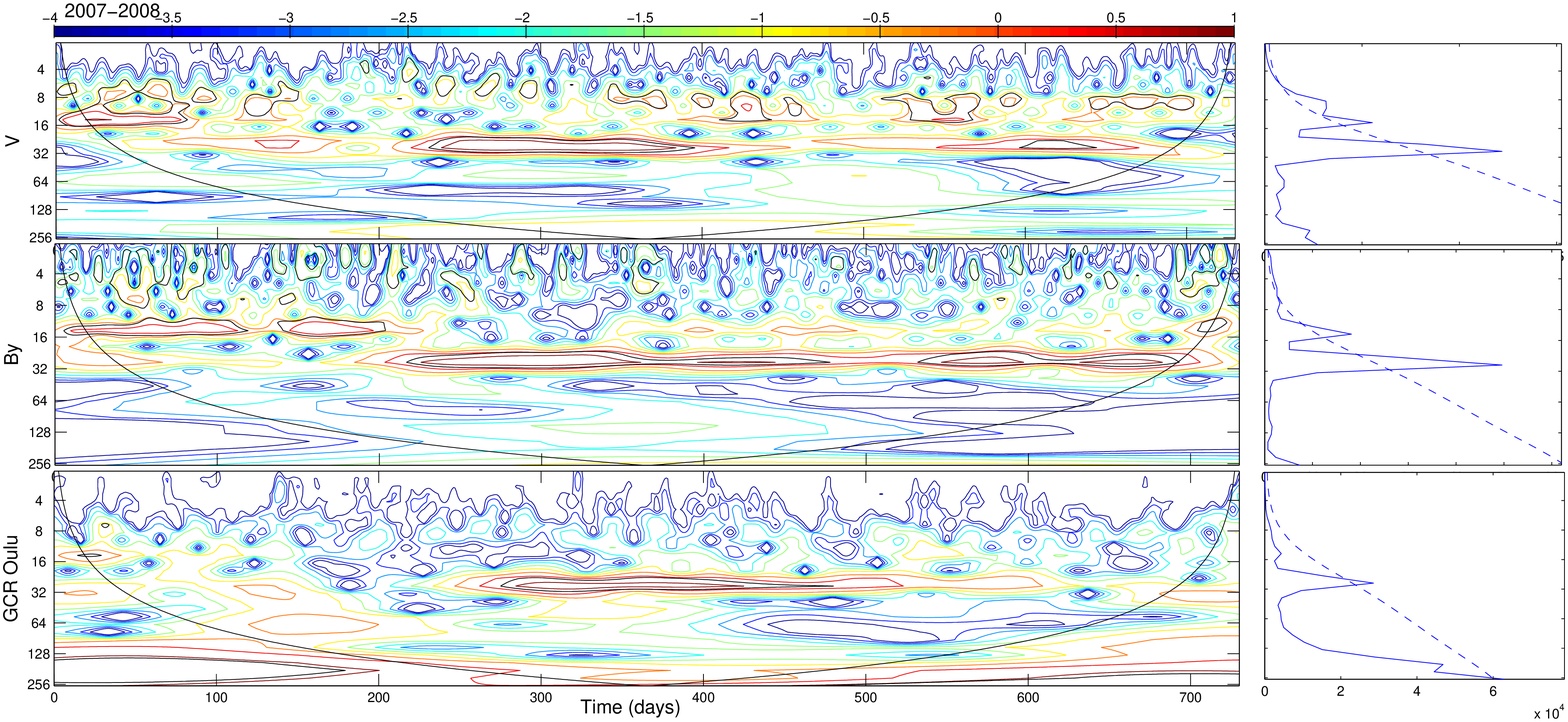}
\end{center}
 \caption{ \label{fig:W20071} Wavelet analysis of the daily solar wind velocity $V$ (a), $By$ component of the IMF (b), GCR intensity for Oulu NM (c) for 2007-2008.}
 \end{figure}
  \begin{figure}[tbp]
  \begin{center}
 \includegraphics[width=1.0\linewidth]{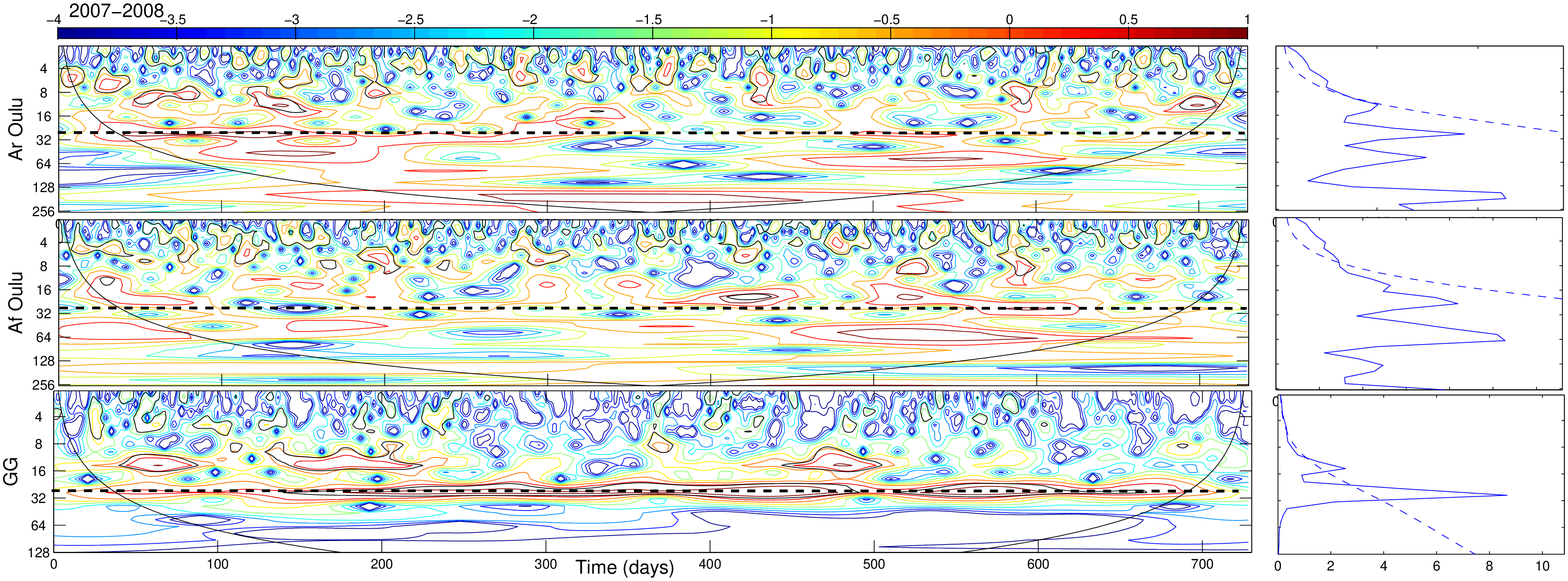}
\end{center}
 \caption{ \label{fig:W20072} Wavelet analysis of the $Ar$ (a) and $Af$ (b) components of the 2D GCR anisotropy and daily $GG$ index (c) for 2007-2008. Dashed line designates the period of 27 days.}
 \end{figure}

  \begin{figure}[tbp]
  \begin{center}
 \includegraphics[width=1.0\linewidth]{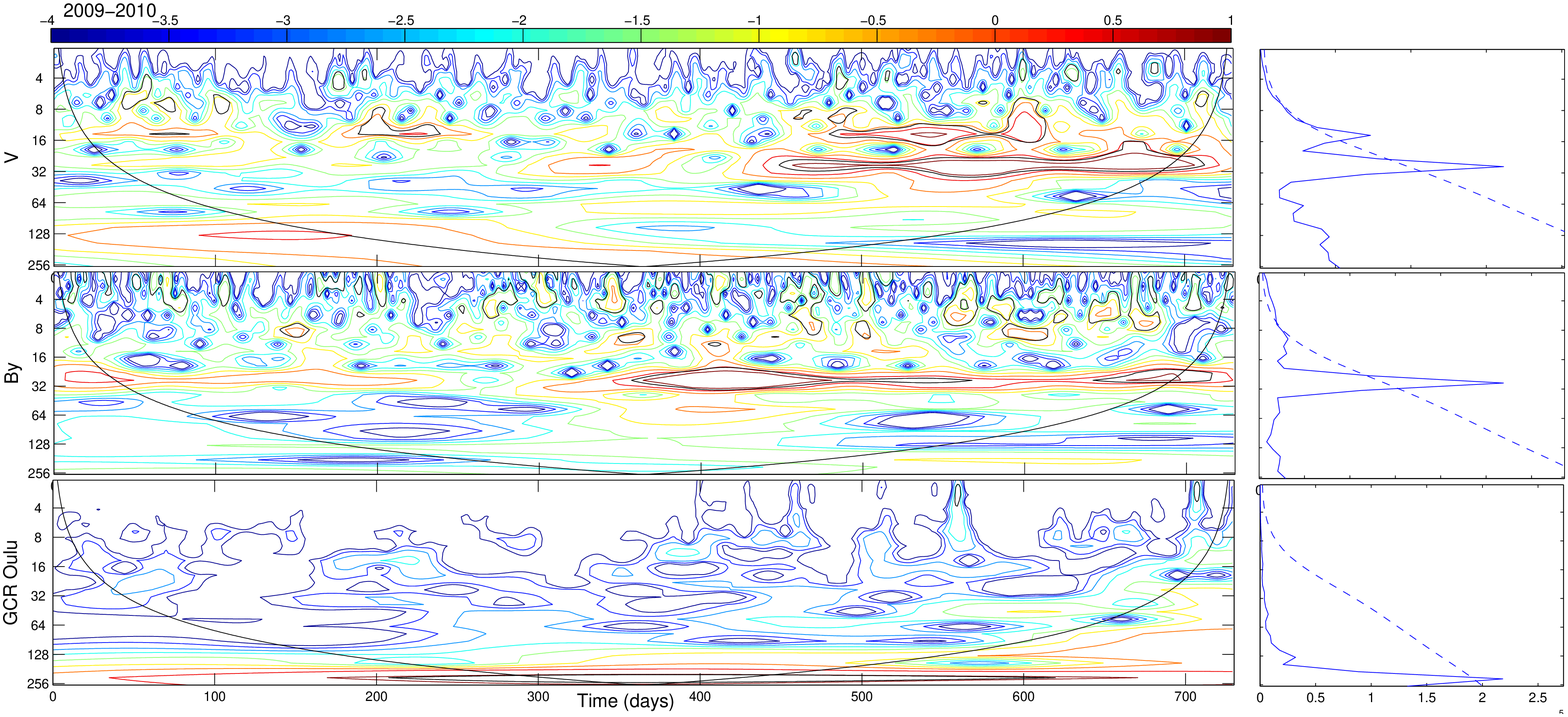}
\end{center}
 \caption{ \label{fig:W20091} Wavelet analysis of the daily solar wind velocity $V$ (a), $By$ component of the IMF (b), GCR intensity for Oulu NM (c) for 2009-2010.}
 \end{figure}
  \begin{figure}[tbp]
  \begin{center}
 \includegraphics[width=1.0\linewidth]{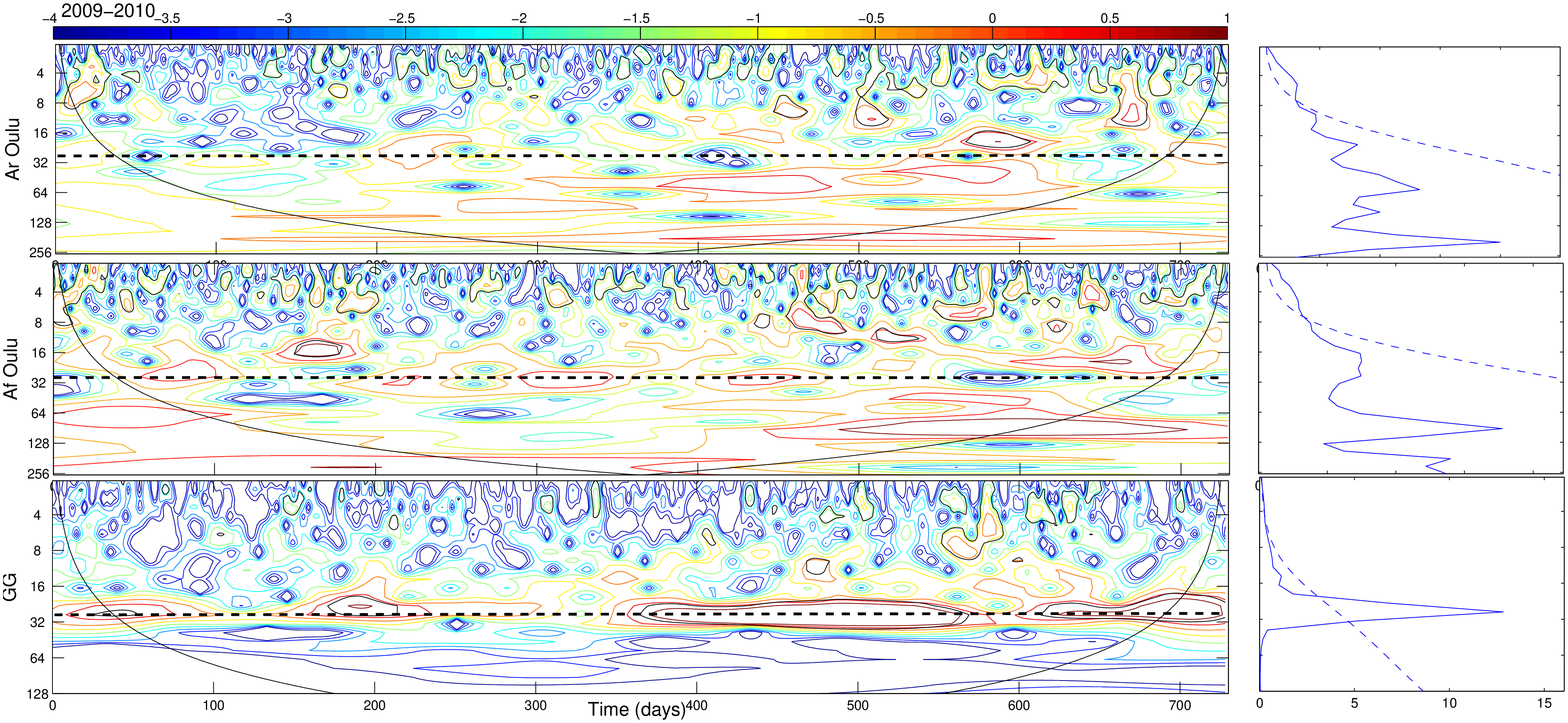}
\end{center}
 \caption{ \label{fig:W20092} Wavelet analysis of the  $Ar$ (a) and $Af$ (b) components of the 2D GCR anisotropy and daily $GG$ index (c) for 2009-2010. Dashed line designates the period of 27 days.}
 \end{figure}
\section{27-day wave in GCR anisotropy near the solar minimum 23/24: 2007-2012}
The 27-day variation of the GCR anisotropy has been studied less intensively up to present, in general.  It partially is connected with the complexity of the reliable revealing of the 27-day variation of the GCR anisotropy by means of small amplitudes of the diurnal variations of GCR ($< 0.3\%$, measured by NMs), and with a large dispersion of the amplitudes of diurnal variation  comparable with the accuracy of hourly data of NMs. In papers Alania et al., \cite{Alania08}, \cite{Alania05} studied the 27-day variation of the 2D GCR anisotropy in the ecliptic plane. They demonstrated  that the average amplitude in the minimum epoch of solar activity is polarity dependent, as it is expected from the drift theory.

The 27-day variation of the north-south anisotropy was studied in the series of papers by Swinson and coauthors \cite{SY92}-\cite{SF95}. They showed that 27-day variation of the north-south anisotropy is correlated with solar activity and this correlation is not clearly dependent upon solar magnetic polarity.

The recent solar minimum provided an unique opportunity to study recurrent variations under relatively stable low solar activity conditions. Recurrent variations connected with corotating structures ($\sim 27$ days) are clearly established in all solar wind and interplanetary parameters. Consequently the 27-day recurrent variations of cosmic ray intensity were clearly seen in a variety of cosmic ray counts detected by neutron monitors (e.g., \cite{Alania10}, \cite{MA13}) and space probes (e.g., \cite{Leske13}).
Recently, Yeeram et al., \cite{Y} studied recurrent 'trains' (trend) of enhanced GCR anisotropy under influence of corotating solar wind structures near the recent solar minimum.

  \begin{figure}[tbp]
  \begin{center}
 \includegraphics[width=1.0\linewidth]{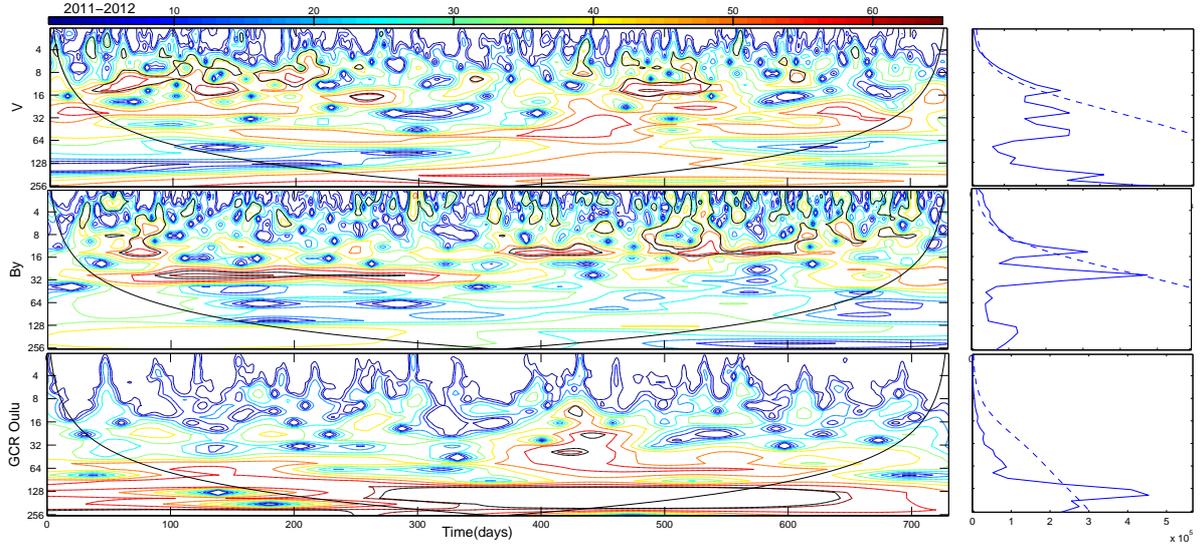}
\end{center}
 \caption{ \label{fig:W20111}Wavelet analysis of the daily solar wind velocity $V$ (a), $By$ component of the IMF (b), GCR intensity for Oulu NM (c) for 2011-2012.}
 \end{figure}
  \begin{figure}[tbp]
  \begin{center}
 \includegraphics[width=1.0\linewidth]{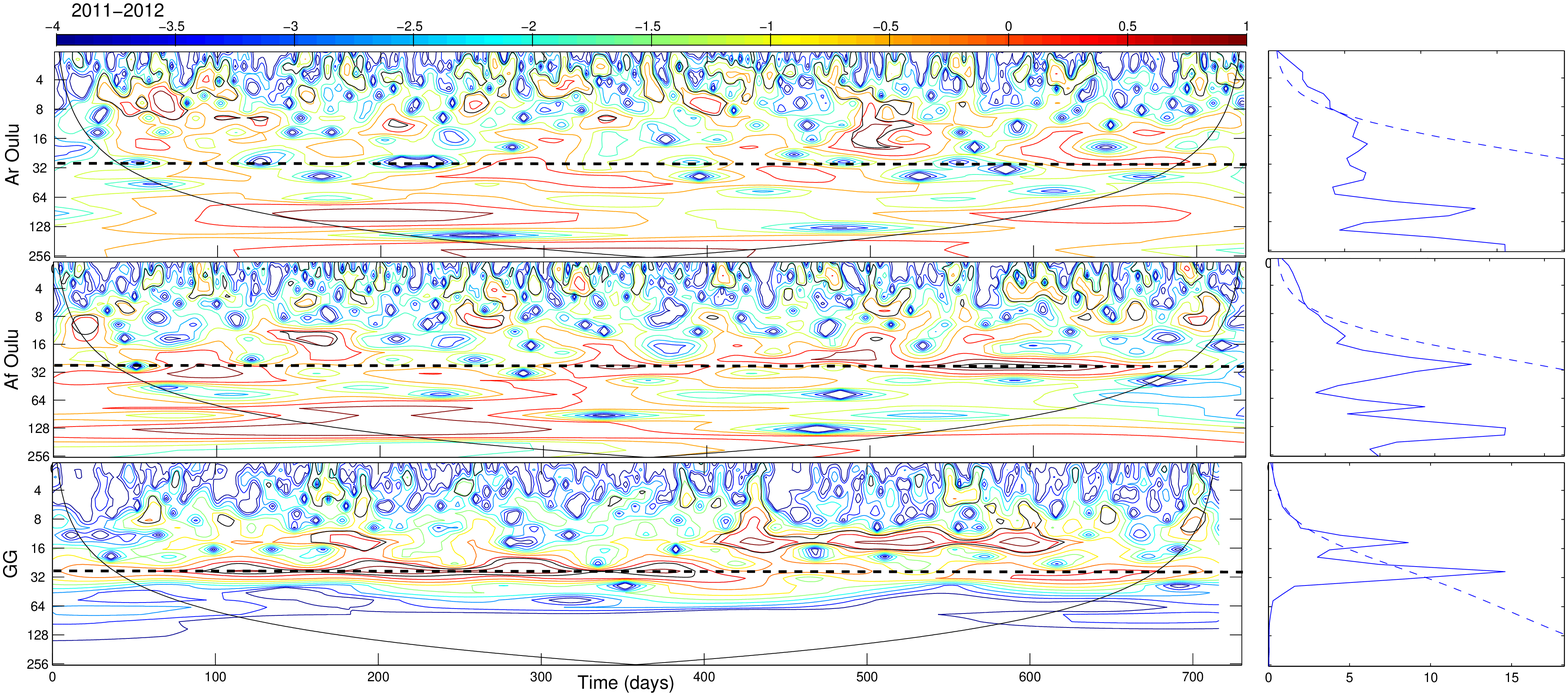}
\end{center}
 \caption{ \label{fig:W20112}Wavelet analysis of the  $Ar$ (a) and $Af$ (b) components of the 2D GCR anisotropy and daily $GG$ index (c) for 2011-2012. Dashed line designates the period of 27 days.}
 \end{figure}

To reveal  the quasi-periodic variation of the 2D GCR  anisotropy in ecliptic plane and  north-south asymmetry of GCR stream by means of $GG$ index near the recent solar minimum 2007-2012, we employ the wavelet time-frequency spectrum technique developed by Torrence and Compo \cite{TC}. The wavelet software is available at the website
[http://paos.colorado.edu/research/wavelets/software.html]. In our calculation we used the Morlet wavelet mother function.

We performed wavelet analysis for the daily solar wind velocity $\mathbf{V}$, $By$ component of the IMF, GCR intensity for Oulu NM, $GG$ index, $Ar$ and $Af$ components of the 2D GCR anisotropy during 2007-2012 considering as a sampling  time interval equaling  two years  (2007-2012). In our case a time interval of two years gives a good enough statistics as far our aim is to reveal  recurrences $\leq$ 27-day period, which is $\sim4\%$ from whole sampling period of 26 solar rotations (two years).

Results of calculation for $\mathbf{V}$, $By$ and GCR intensity are presented in Fig. ~\ref{fig:W20071}  for 2007-2008, Fig. ~\ref{fig:W20091} for 2009-2010, Fig. ~\ref{fig:W20111} for 2011-2012,  respectively. Wavelet analysis of $Ar$ and $Af$ components of the 2D GCR anisotropy for Oulu NM and $GG$ index are presented in Figs. ~\ref{fig:W20072} for 2007-2008, Fig. ~\ref{fig:W20092} for 2009-2010, Fig.  ~\ref{fig:W20112} for 2011-2012,  respectively.
Figures ~\ref{fig:W20072}, ~\ref{fig:W20092}, ~\ref{fig:W20112} panel (c) present very clear quasi-periodic changes in $GG$ index related to the Sun's rotation ($\sim 27$ days) for almost whole time interval 2007-2012. Similar quasi periodic character is clearly visible in $\mathbf{V}$ and $By$ component of the IMF. 27-day variation of GCR intensity is well established at the end of 2007 and in 2008. Although recurrent variations connected with corotating structures ($\sim 27$ days) are clearly established in almost all solar wind parameters, 2D GCR anisotropy  shows a weak 27-day variation, only in some periods. This is connected with large dispersion of daily $Ar$ and $Af$ components of the 2D GCR anisotropy.

\section{Annual waves in $GG$ index near the solar minimum 23/24: 2006-2012}
\begin{figure}[tbp]
  \begin{center}
 \includegraphics[width=0.65\linewidth]{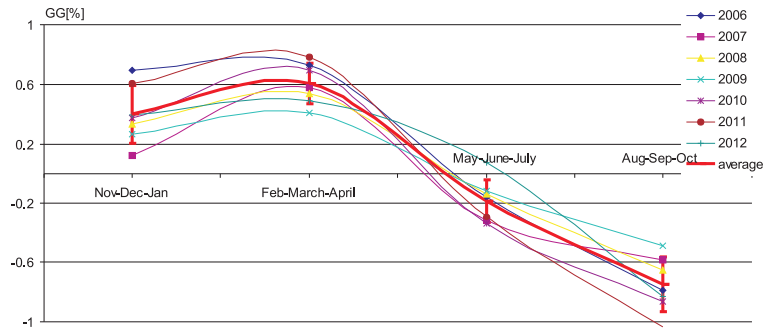}
\end{center}
 \caption{ \label{fig:annual}Superposition of the 3 month average $GG$ index for each year during 2006-2012}
 \end{figure}

We have also studied a feature of the north-south  asymmetry based on the GG index     depending  on the  position of the Earth ($\pm7.5^{0}$ in September and March) during its annual motion  around the Sun. For this purpose  we calculated 3 month average of the $GG$ index  for each year during 2006-2012. In each year 1st point is the average of November-December-January, 2nd point is the average of February-March-April, 3rd point is the average of May-June-July  and 4th point is the average of August-September-October. Results of calculations for superimposed $GG$ index are presented in Fig. ~\ref{fig:annual}. Figure ~\ref{fig:annual}  shows that $GG$ index exhibits clear annual wave with evident maximum phase near the February-March-April period. We suppose that it is an indication of the asymmetry of a heliolatitudinal distribution of solar activity having minimum in south  hemisphere near ~$7.5^{0}$; being some source of north-south asymmetry of the GCR flux in 2007-2012.
Unfortunately, we could not make a precise quantitative estimation due to some uncertainty character of $GG$ data  used in this paper.  At the same time, one can suppose that this finding is an indication of an existence of north-south asymmetry of the GCR flux.

\section{Conclusions}
\begin{enumerate}
  \item We study recurrent variations of the GCR anisotropy in the ecliptic plane and in the polar direction connected with corotating structures observed in the heliosphere near the recent solar minimum in the period 2007-2012.
  \item 2D GCR anisotropy generally does not show a clear  evidence of the 27-day variation, but in some periods weak recurrent character is visible. This is partially connected with large dispersion of daily $Ar$ and $Af$ components of the 2D GCR anisotropy.
  \item Using wavelet time-frequency method we reveal clear 27-day waves in the $GG$ index, being the measure of the north-south anisotropy, for almost whole analyzed period 2007-2012. Similar recurrent properties are observed in  solar wind velocity and interplanetary magnetic field.
  \item We show that $GG$ index exhibits a clear annual wave with an evident phase of maximum near the February-March-April period. One can suppose that  possibly it is an indication of the asymmetry of a heliolatitudinal distribution of solar activity having minimum in south  hemisphere near $\sim 7.5^{0}$; this heliolatitudinal asymmetry can be considered as a  some source of north-south asymmetry of the GCR flux in 2007-2012.

\end{enumerate}

\section*{Acknowledgments}
 We thank the principal investigators of Oulu neutron monitor and Nagoya muon telescope for the ability to use their data. We acknowledge  authors of the wavelet software which is available at URL: http://paos.colorado.edu/research/wavelets/.
 We would like to thank the reviewers for helpful suggestions.

\end{document}